\documentclass[conference]{IEEEtran}
\IEEEoverridecommandlockouts
\usepackage[hidelinks]{hyperref}
\usepackage{amsmath,amssymb,amsfonts}
\usepackage{algorithmic}
\usepackage{graphicx}
\usepackage{textcomp}
\usepackage{xcolor}
\usepackage{siunitx}
\usepackage{subcaption}
\usepackage{multirow}
\usepackage{atbegshi}
\usepackage{xsavebox}

\hypersetup{
    pdfauthor={Paul-Otto Müller and Oskar von Stryk},
    pdftitle={The Foundation for Developing an Exoskeleton for the Rehabilitation of Temporomandibular Disorders},
    pdfsubject={Temporomandibular disorders affect a significant portion of the population, often causing pain and restricted jaw movement. Physiotherapy with active jaw exoskeletons seems to be a promising treatment approach. However, there has been limited progress in this area, with only a few studies published. The lack of detailed simulations and biomechanical jaw models in these papers impedes the design and validation of such devices. Thus, this work presents an open-source and extendable jaw model framework to support the development of jaw exoskeletons for TMD rehabilitation. The framework includes three model variants with varying complexity: a rigid body model with point-on-surface constraints, a rigid body model with mesh geometry constraints, and a hybrid model combining rigid dynamics with finite element methods. Model customization is facilitated through \textit{JSON} configuration files, requiring minimal programming expertise. The framework was evaluated using experimental kinematic data from a single participant, with muscle excitations computed via PID controllers tuned through global optimization. The models successfully reproduced jaw movements without prior individualization, demonstrating their suitability for prototyping and studying exoskeleton behavior. The basic models are ideal for rapid prototyping, while the hybrid model provides insights into the effects of exoskeletons on the masticatory system. This framework establishes a foundation for advancing jaw exoskeleton development, with future work focusing on patient-specific modeling, improved anatomical detail, and tools for exoskeleton control and design. The framework code is publicly available and can be found at \href{https://github.com/paulotto/exosim}{https://github.com/paulotto/exosim}.},
    pdfkeywords={exoskeletons, jaw model, temporomandibular disorders, rehabilitation, framework}
}

\def\BibTeX{{\rm B\kern-.05em{\sc i\kern-.025em b}\kern-.08em
    T\kern-.1667em\lower.7ex\hbox{E}\kern-.125emX}}

\newcommand{\etal}{et~al\@.}

\makeatletter
\newcommand{\Autoref}[1]{\@first@ref#1,@}
\def\@throw@dot#1.#2@{#1}%
\def\@set@refname#1{%
    \edef\@tmp{\getrefbykeydefault{#1}{anchor}{}}%
    \xdef\@tmp{\expandafter\@throw@dot\@tmp.@}%
    \ltx@IfUndefined{\@tmp autorefnameplural}%
         {\def\@refname{\@nameuse{\@tmp autorefname}s}}%
         {\def\@refname{\@nameuse{\@tmp autorefnameplural}}}%
}
\def\@first@ref#1,#2{%
  \ifx#2@\autoref{#1}\let\@nextref\@gobble%
  \else%
    \@set@refname{#1}%
    \@refname~\ref{#1}%
    \let\@nextref\@next@ref%
  \fi%
  \@nextref#2%
}
\def\@next@ref#1,#2{%
   \ifx#2@ and~\ref{#1}\let\@nextref\@gobble%
   \else, \ref{#1}%
   \fi%
   \@nextref#2%
}
\makeatother
    
\begin{document}

\title{The Foundation for Developing an Exoskeleton for the Rehabilitation of Temporomandibular Disorders\\
}

\author{\IEEEauthorblockN{1\textsuperscript{st} Paul-Otto Müller}
\IEEEauthorblockA{\textit{Simulations, Systems Optimization and Robotics Group} \\
\textit{Technical University of Darmstadt}\\
Darmstadt, Germany \\ 
pmueller@sim.tu-darmstadt.de}
\and
\IEEEauthorblockN{2\textsuperscript{nd} Oskar von Stryk}
\IEEEauthorblockA{\textit{Simulations, Systems Optimization and Robotics Group} \\
\textit{Technical University of Darmstadt}\\
Darmstadt, Germany \\
stryk@sim.tu-darmstadt.de}
}

\makeatletter
\newcommand\notsotiny{\@setfontsize\notsotiny{6.6}{7.7}}
\makeatother

\xsavebox{PageTopWatermark}{%
    \begin{minipage}{\paperwidth}
        \notsotiny
        \centering
        Preprint of the paper which appeared in: 2025 IEEE INTERNATIONAL CONFERENCE ON SIMULATION, MODELING, AND PROGRAMMING FOR AUTONOMOUS ROBOTS (SIMPAR).
    \end{minipage}
}

\xsavebox{PageBottomWatermark}{%
    \begin{minipage}{\paperwidth}
        \notsotiny
        \centering
        \parbox{18cm}{\centering
        \textcopyright 2025 IEEE. Personal use of this material is permitted. Permission from IEEE must be obtained for all other uses, in any current or future media, including reprinting/republishing
        this material for advertising or promotional purposes, creating new collective works, for resale or redistribution to servers or lists, or reuse of any copyrighted component of this work in other works.}
    \end{minipage}
}

\AtBeginShipout{
    \AtBeginShipoutUpperLeft{\raisebox{-0.8cm}{\xusebox{PageTopWatermark}}}
}

\AtBeginShipout{
    \AtBeginShipoutUpperLeft{\raisebox{-27.3cm}{\xusebox{PageBottomWatermark}}}
}

\maketitle

\begin{abstract}
Temporomandibular disorders affect a significant portion of the population, often causing pain and restricted jaw movement. Physiotherapy with active jaw exoskeletons seems to be a promising treatment approach. However, there has been limited progress in this area, with only a few studies published. The lack of detailed simulations and biomechanical jaw models in these papers impedes the design and validation of such devices. Thus, this work presents an open-source and extendable jaw model framework to support the development of jaw exoskeletons for TMD rehabilitation. The framework includes three model variants with varying complexity: a rigid body model with point-on-surface constraints, a rigid body model with mesh geometry constraints, and a hybrid model combining rigid dynamics with finite element methods. Model customization is facilitated through \textit{JSON} configuration files, requiring minimal programming expertise. The framework was evaluated using experimental kinematic data from a single participant, with muscle excitations computed via PID controllers tuned through global optimization. The models successfully reproduced jaw movements without prior individualization, demonstrating their suitability for prototyping and studying exoskeleton behavior. The basic models are ideal for rapid prototyping, while the hybrid model provides insights into the effects of exoskeletons on the masticatory system. This framework establishes a foundation for advancing jaw exoskeleton development, with future work focusing on patient-specific modeling, improved anatomical detail, and tools for exoskeleton control and design. The framework code is publicly available and can be found at \href{https://github.com/paulotto/exosim}{https://github.com/paulotto/exosim}.
\end{abstract}


\section{Introduction}\label{sec:introduction}

Temporomandibular disorders (TMDs) are commonly known as diseases and issues that originate from the masticatory system and the temporomandibular joints (TMJs), affecting approximately \SI{10}{\%} of the global population \cite{Tran2022}. The masticatory system, one of the most complex systems in the human body and responsible for jaw movement involved in chewing, swallowing, and speaking, is prone to various disorders. These disorders are typically expressed through symptoms such as pain, restricted jaw movement, clicking or popping sounds, and muscle stiffness \cite{Durham2015}. In severe cases, conditions such as arthritis or joint dislocations may also be observed. Noninvasive rehabilitation approaches, particularly physiotherapy, are widely recognized as the first-line treatment for TMDs, as they are effective in most cases \cite{Shimada2019}. Surgical interventions are generally considered only when conservative treatments fail. For rehabilitation, an active jaw exoskeleton might be a promising tool, as it could provide support during physical training, assist therapists, and introduce new possibilities for adaptable training modalities, neurological feedback, and progress monitoring \cite{Xie2016}. Additionally, jaw exoskeletons could support patients recovering from accidents or TMJ surgeries.

Developing a jaw exoskeleton for TMD rehabilitation involves addressing several scientific challenges. Among these are ensuring user safety, acceptance, wearability, and effective sensor and actuator integration. The biomechanical jaw model is central to the design of such an exoskeleton, which enables iterative prototyping, testing, and validation of the integrity of the design. A model is the primary method to analyze the force conditions within the TMJs and facilitates comparisons between healthy and pathological jaw movements. Depending on the specific task, jaw models of varying levels of detail and complexity are required. For instance, a basic, real-time-capable model may be sufficient for control purposes. In contrast, a more detailed model is necessary for assessing the effects of the exoskeleton on the masticatory system and TMJs. The dynamics of mastication, characterized by a combination of rotational and translational movements following an S-shaped trajectory during jaw opening, demand a detailed and accurate representation \cite{Wilkie2022}.

To address these requirements, an extendable open-source jaw model framework is introduced. This framework includes models of varying complexity and detail, intended to serve as the foundation for developing a jaw exoskeleton for TMD rehabilitation, a field still in its early stages. The framework abstracts model components, properties, and parameters into \textit{JSON} files, allowing for easy customization without requiring advanced programming skills. Furthermore, a novel framework feature includes a model that connects the superior head of the lateral pterygoid muscles to the TMJ discs, an aspect rarely implemented in existing literature.

The open-source framework is not presented as a comprehensive or definitive jaw model collection but rather as a foundational tool for developing jaw exoskeletons designed for TMD treatment, which remains underexplored in the scientific literature. By facilitating prototyping and enabling the study of exoskeleton effects on the masticatory system, the framework aims to advance research in this emerging field. Its open-source nature encourages research collaboration, ensuring a shared foundation for further investigation and innovation.

\section{Related Work}\label{sec:related-work}

\subsection{Jaw Exoskeletons}

Research on exoskeletons for treating TMDs is still in its early stages, with only a few studies published in this field. A system comprising two rigid four-bar linkage mechanisms, attached to a helmet and connected to the jaw via a chin holder, was developed by Wang \etal{} \cite{Wang2010, Wang2014}. This system was designed to support or resist opening and closing motions and was controlled using a manually tuned PID controller. However, only a passive, rigid jaw model---neither based on nor validated with experimental data---was employed to test the device. 

A concept for a shoulder-mounted exoskeleton intended to assist jaw movements in the vertical direction was presented by Evans \etal{} \cite{Evans2016}. The system consisted of a rigid, rotating bracket with an attached mouthpiece and chin strap. However, the device was neither simulated nor prototyped. A semi-powered exoskeleton for strengthening oral motor function was developed by Kameda \etal{} \cite{Kameda2021}. This system included a chin cup connected to the head by cables, which introduced inherent compliance, and shape-memory alloy springs to assist the user in closing the jaw. The device was tested on ten healthy subjects, but no jaw model was used to validate its functionality.

The first soft approach to a jaw exoskeleton was introduced by Zhang \etal{}, who developed a system using two pneumatic actuators attached to the head and chin with adjustable straps \cite{Zhang2021}. This system focused primarily on actuator design; no virtual jaw model was utilized for design or validation.

Although these studies provide valuable insights and requirements for designing jaw exoskeletons, they lack the integration of a detailed biomechanical jaw model. Such a model is essential for enabling rapid prototyping, testing, and validation of jaw exoskeleton systems.

\subsection{Biomechanical Jaw Models}

The development of biomechanical jaw models has a long history and is supported by an extensive body of related work. These models are generally categorized as either static or dynamic, as well as rigid, finite element models (FEM), or hybrid models that combine both approaches. Early examples of three-dimensional dynamic models of the human masticatory system were introduced by Weingartner \etal{} and Koolstra and van Eijden \cite{Koolstra1997, Weingartner1997}. In the former, revolute joints and circular shapes were used to approximate the path of the condyles, while in the latter, the condyles were modeled as spheres capable of contacting the articular fossae, which were represented as polynomial functions. Additionally, force-length and force-velocity dependencies of the muscles were integrated into Koolstra and van Eijden's model. Peck \etal{} adopted a similar approach, representing the condyles as ellipsoids that slide frictionlessly on the articular fossae \cite{Peck2000}. 

The TMJs have been modeled with point-on-plane constraints in the works of Stavness \etal{} and Zee \etal{}, where the movement of the condyles was restricted to predefined paths \cite{Stavness2006, Zee2007}. Andersen \etal{} proposed a force-dependent approach, allowing the condyles to move freely within the fossae while accounting for joint geometry and elasticity \cite{SkipperAndersen2017}. A model partly driven by electromyography data and employing forward and inverse dynamics simulations to predict mandibular movement was developed by Guo \etal{}, with the TMJs modeled as spherical elements interacting with a frictionless surface \cite{Guo2022}.

FEMs are more computationally expensive but provide detailed insights into stress and strain distributions within the jaw. These models often include the mandible, TMJ discs, surrounding muscles, and even teeth \cite{Kober2004, MartinezChoy2017, Vukicevic2021, Gholamalizadeh2022}.

Hybrid models aim to combine the advantages of rigid and FEM approaches. An early implementation by Koolstra and van Eijden incorporated FEM representations for TMJ discs and cartilage while treating the bony structures as rigid \cite{Koolstra2005}. More recent examples, such as the models by Sagl \etal{}, have extended this approach to include FEM representations of the TMJ capsules as well \cite{Sagl2019a, Sagl2021}.

Data-driven models also exist, such as the kinematic model by Yang \etal{}, which employs nonlinear principal component analysis to generate realistic jaw movements \cite{Yang2019}.

Despite these advancements, many models remain closed-source and lack extensibility or customizability. An open-source approach would be highly beneficial for fostering collaboration and accelerating progress in this field.

\section{A Jaw Model Framework}\label{sec:jaw-model-framework}

The multi-body physics engine \textit{Project Chrono} was selected as the simulation engine for the jaw model framework \cite{Tasora2016}. This choice was motivated by its open-source nature, active community, ongoing maintenance, support for both FEM and rigid body dynamics, and the ability to extend the engine with custom components. By abstracting model components, properties, and parameters into \textit{JSON} files, the proposed jaw model framework allows for easy customization without requiring extensive programming expertise.

\subsection{Requirements}

The degree of detail and complexity required for the jaw model depends on its intended application. A basic and fast model is necessary for the design, rapid prototyping, and integrity checks of a jaw exoskeleton. A more complex and realistic model is required to compare healthy individuals with TMD-affected patients or estimate the effects of the exoskeleton on the masticatory system and joints. A real-time capable model is also essential for designing and controlling the exoskeleton.

The framework was designed to be extendable and customizable. Public availability is prioritized to encourage collaboration and accelerate progress by enabling researchers to build upon a shared foundation, particularly in a research field still in its early stages.

\subsection{Model Structure}

\begin{figure}[htbp]
    \centering
    \begin{subfigure}[]{0.28\textwidth} %
        \centering
        \includegraphics[width=\textwidth]{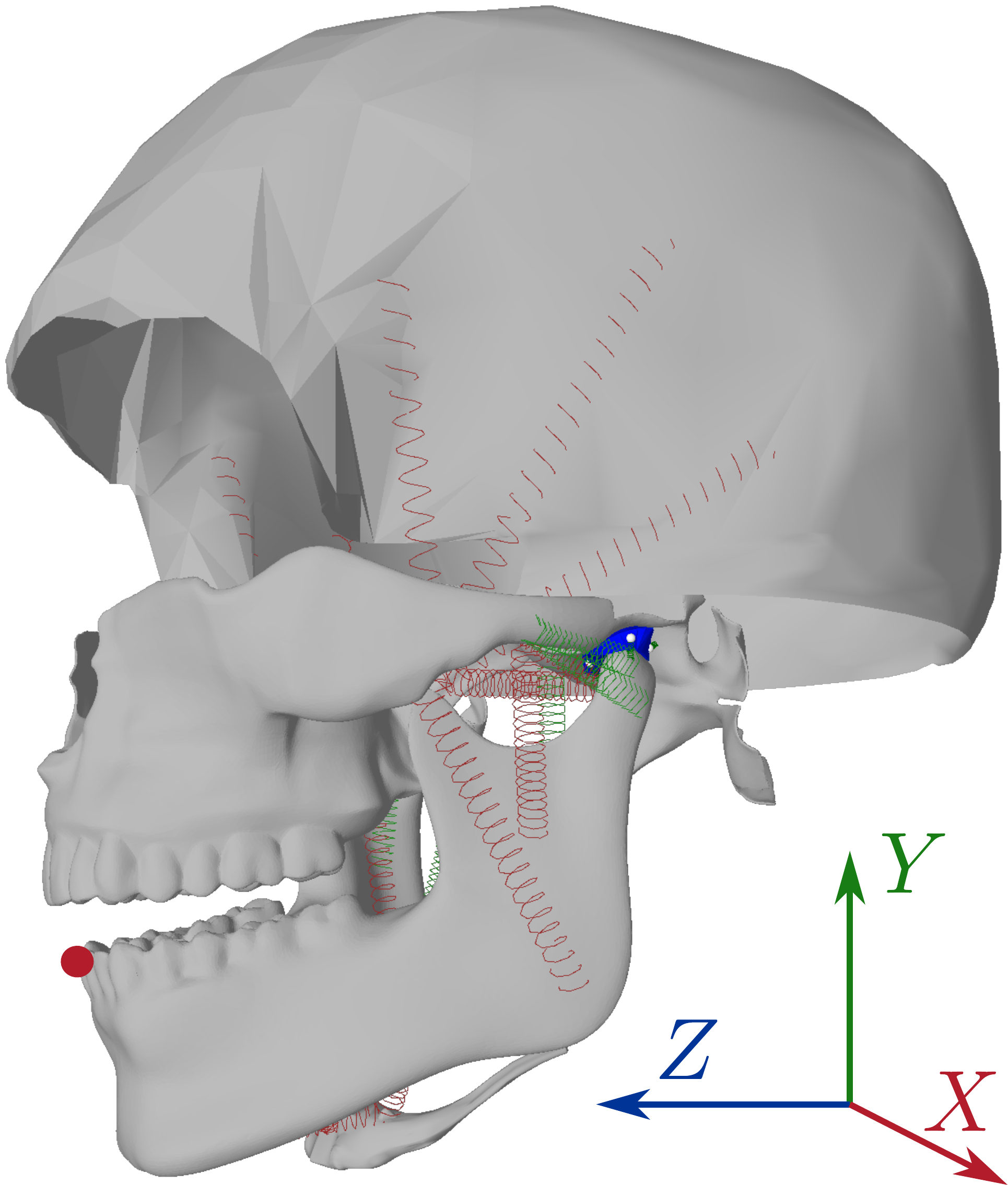}
        \caption{} 
        \label{fig:jaw_model_fem_trans}
    \end{subfigure}
    \hspace{6mm}
    \begin{subfigure}[]{0.11\textwidth} %
        \centering
        \includegraphics[width=\textwidth]{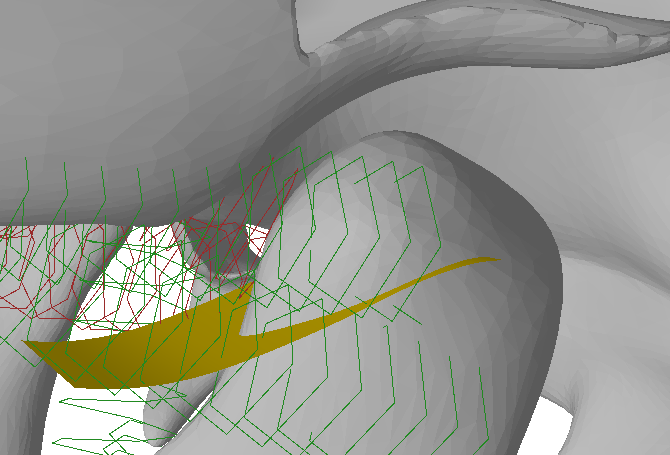}
        \caption{}
        \label{fig:tmj_point_on_surface}
        
        \includegraphics[width=\textwidth]{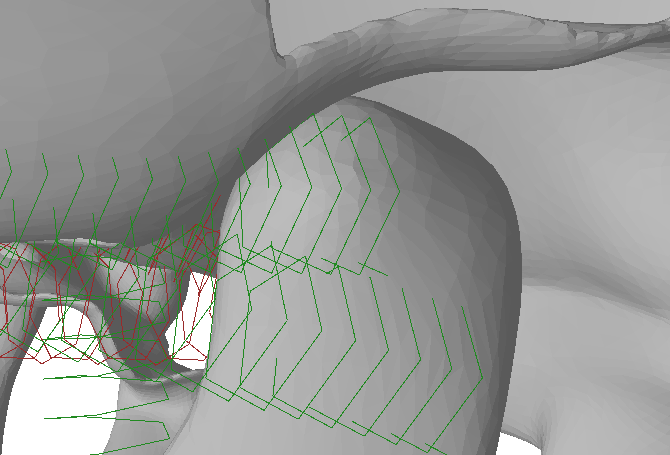} 
        \caption{}
        \label{fig:tmj_mesh_geometries}
        
        \includegraphics[width=\textwidth]{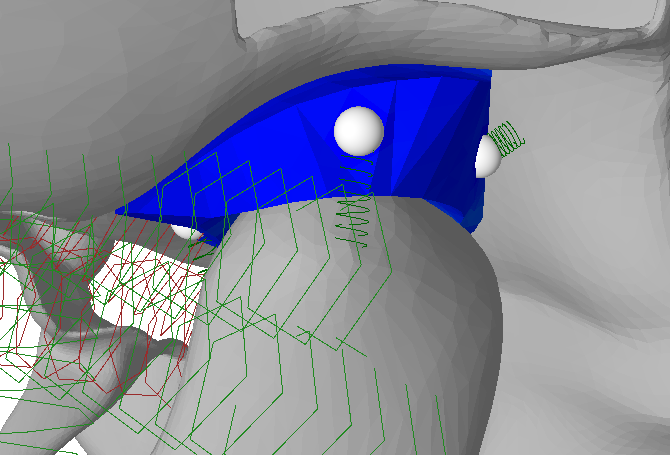}
        \caption{} 
        \label{fig:tmj_fem}
    \end{subfigure}
    \captionsetup{font=small, labelfont=bf}
    \caption{(a) The hybrid jaw model (red: muscles, green: ligaments, blue: TMJ disc, gray: bones). The red dot in the middle of the lower teeth marks the incisal point. 
    (b) Point-on-surface constraint (yellow). (c) Mesh geometry contact constraint. (d) FEM disc.}
    \label{fig:jaw_model}
\end{figure}

The framework includes three jaw model variants, each with a different approach to modeling the TMJs and varying levels of complexity, depending on the specific use case.

1. Rigid Body Model with Point-on-Surface Constraints: This model constrains the motion of the mandibular condyles to a user-defined surface (\autoref{fig:tmj_point_on_surface}), approximating the anatomical path dictated by the articular fossa and eminence. A set of control points defines the surface.

2. Rigid Body Model with Natural Geometries: In this approach, the condyles' motion is bounded by contact between the natural geometries derived from bone meshes, resulting in a more anatomically accurate movement (\autoref{fig:tmj_mesh_geometries}).

3. Hybrid Model with FEM Discs: This model combines rigid body dynamics with FEM to represent the viscoelastic TMJ discs situated between the condyles and fossae (\Autoref{fig:jaw_model_fem_trans, fig:tmj_fem}). The TMJ discs absorb and distribute joint forces and are connected by ligaments to the articular fossae and condyles in the model. This model is based on the work of Sagl \etal{} \cite{Sagl2019a}. A novel feature of this framework is the inclusion of a connection between the superior head of the lateral pterygoid muscles and the TMJ discs, counteracting disc retraction during mandible closing and posterior shifting of the condyles. This feature is anatomically suggested but has not been previously implemented in the literature \cite{Okeson2019}.

All models include the mandible, maxilla, hyoid, and cranium, which are represented as rigid bodies \cite{Hannam2008}. The system is driven by 24 muscles attached to bones via tendons. Various muscle models, such as those by Millard \etal{} and Peck \etal{}, are implemented \cite{Millard2013, Peck2000}. Eight ligaments, based on the model by Blankevoort and Huiskes, are included to stabilize and constrain jaw motion \cite{Blankevoort1991}. Parameters were chosen according to the literature.  

\textit{Blender} was utilized to modify and process the bone meshes, create collision meshes, and define attachment points for muscles and ligaments. TMJ discs were generated by taking the imprint between mandibular condyles and fossae. The resulting disc surface meshes were converted into tetrahedral volume meshes using \textit{fTetWild} \cite{Hu2020}.

The framework allows for arbitrary increases in model complexity, as all components can be replaced with more advanced FEM representations.

\subsection{Evaluation of the Model Framework}

To evaluate the jaw model framework, an experiment with one participant was conducted, and kinematic data was collected using a \textit{Qualisys Oqus} motion capture system recording at \SI{200}{Hz}. The study followed the ethical principles outlined in the Declaration of Helsinki. Ethical approval was obtained from the ethics committee of TU Darmstadt (approval number EK \num{03}/\num{2025}), and the participant provided written informed consent prior to their inclusion in the study.

\begin{figure}
    \centering
    \includegraphics[width=0.45\textwidth]{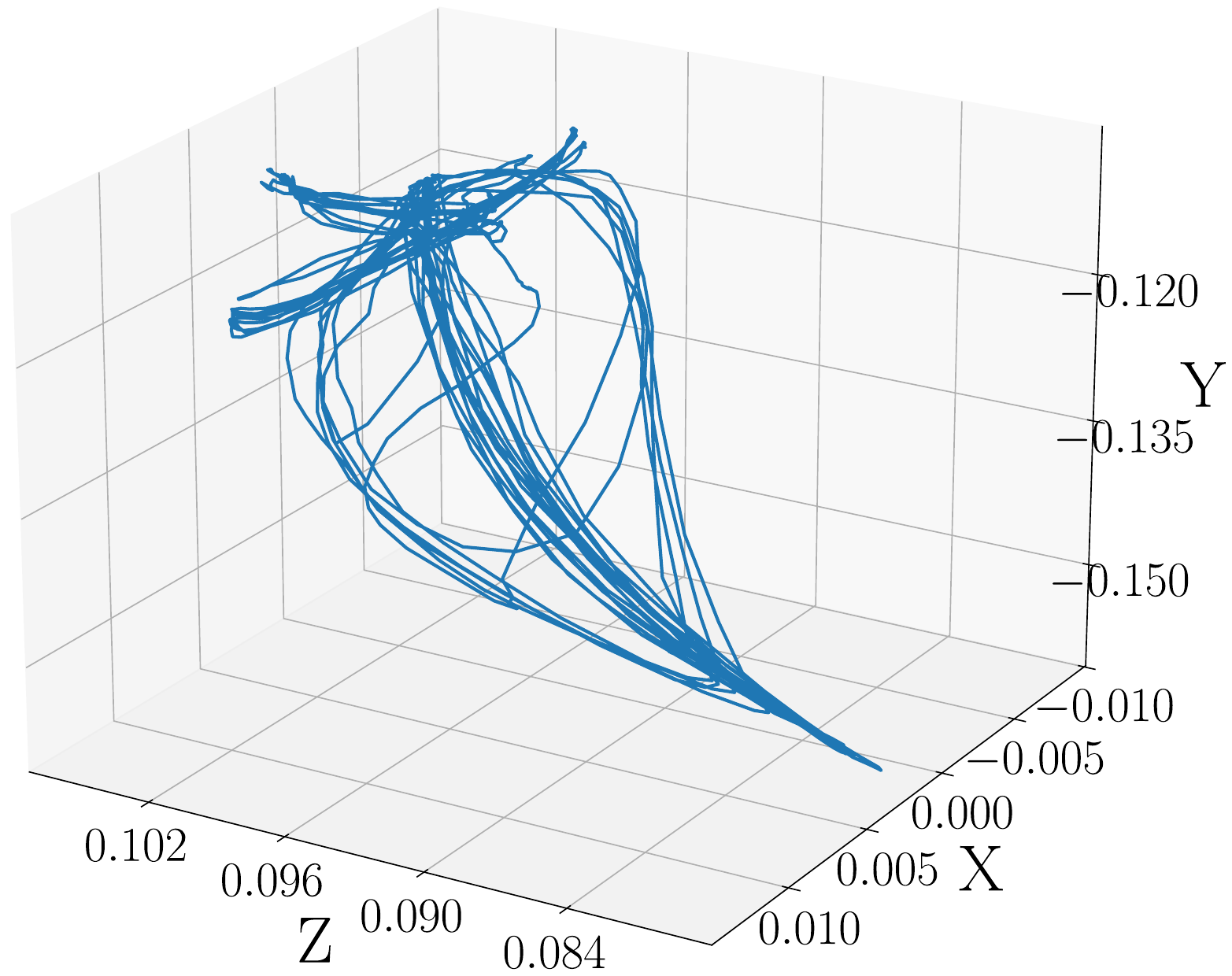}
    \captionsetup{font=small, labelfont=bf}
    \caption{A 3D plot of all recorded jaw movements, including opening and closing, protrusion and retrusion, lateral motions, and chewing. The reference point on the jaw was the incisal point in the middle of the lower teeth. The unit is meters.}
    \label{fig:exp-3d-plot}
\end{figure}

Jaw movements were recorded using a 3D-printed mouthpiece with reflective markers, which was temporarily attached to the participant's lower teeth using dental glue. This setup minimized soft tissue artifacts. A marker headpiece was also placed on the participant's head to account for head motion in the jaw movement data. Three points were marked on the upper and lower teeth using a 3D-printed calibration object with known dimensions to calculate a relative transformation between the motion capture system and the virtual model. These points were identified in both the physical and virtual systems, eliminating the need to know the exact relative pose of the mouthpiece to the mandible. The participant was instructed to perform a series of jaw movements, including opening and closing, protrusion and retrusion, lateral motions, and chewing. A plot of all the recorded trajectories is shown in \autoref{fig:exp-3d-plot}.

The experimental data was used to demonstrate that the models within the framework can be template models capable of reproducing jaw movements from a real individual. This was achieved without prior personalization of the model. However, individualization of the model would eventually be necessary for more accurate predictions tailored to a specific TMD patient.

The muscle excitations required to reproduce the recorded jaw movements were determined through a two-step process, with each muscle controlled by a PID controller. In the first step, the target muscle lengths for the controllers were obtained by aligning the mandible mesh to the recorded poses, satisfying model constraints by solving a system of nonlinear equations, and extracting the static muscle lengths. The controller parameters were optimized in the second step by minimizing the objective function
\begin{equation*}
    J = \sum_{i=0}^{N} \left[ \alpha \| \mathbf{q}_{\text{observed},i} - \mathbf{q}_{\text{simulated},i} \|^2_2 + \beta \| \mathbf{u}_i \|^2_2 + \gamma \| \dot{\mathbf{u}}_i \|^2_2 \right], 
\end{equation*}%
where $\mathbf{q}_{\text{observed,i}}$ and $\mathbf{q}_{\text{simulated,i}}$ are the observed and simulated mandible poses at time step $i$ with $N$ data points, respectively. This objective focuses on the global position error rather than the local muscle lengths. The muscle excitations of all muscles $\mathbf{u}$ and their derivatives $\dot{\mathbf{u}}$ are included to penalize high muscle activations and rapid changes in muscle excitations. The weighting factors are set to $\alpha = \num{1}$, $\beta = \num{0.1}$, and $\gamma = \num{0.1}$. Note that the objective terms were scaled before multiplying the weights to ensure that the terms were of similar magnitudes.
The dynamic optimization problem, involving 72 dimensions, was solved using the \textit{Pagmo} optimization library and a global self-adaptive differential evolution algorithm \cite{Biscani2020}. By directly applying the muscle excitations generated by the controllers and simulating the system, dynamic effects were accounted for, even though the muscle lengths could also be set directly.

The optimization process was conducted on a high-performance computer. Simulations with the PID controllers were run on a laptop equipped with an Intel i\num{7}-\num{13700}HX (\num{24} cores) processor operating at \SI{4.8}{GHz}, \SI{32}{GB} of memory, and running Ubuntu \num{22}.\num{04}.\num{5} LTS. A simulation time step of \SI{1}{ms} was used for all models. The point-on-surface constrained model can run in real-time, while the mesh geometry constrained and hybrid models run approximately \num{40} and \num{2000} times slower than real-time, respectively.

\subsection{Results}

\begin{figure}
    \centering
    \includegraphics[width=0.45\textwidth]{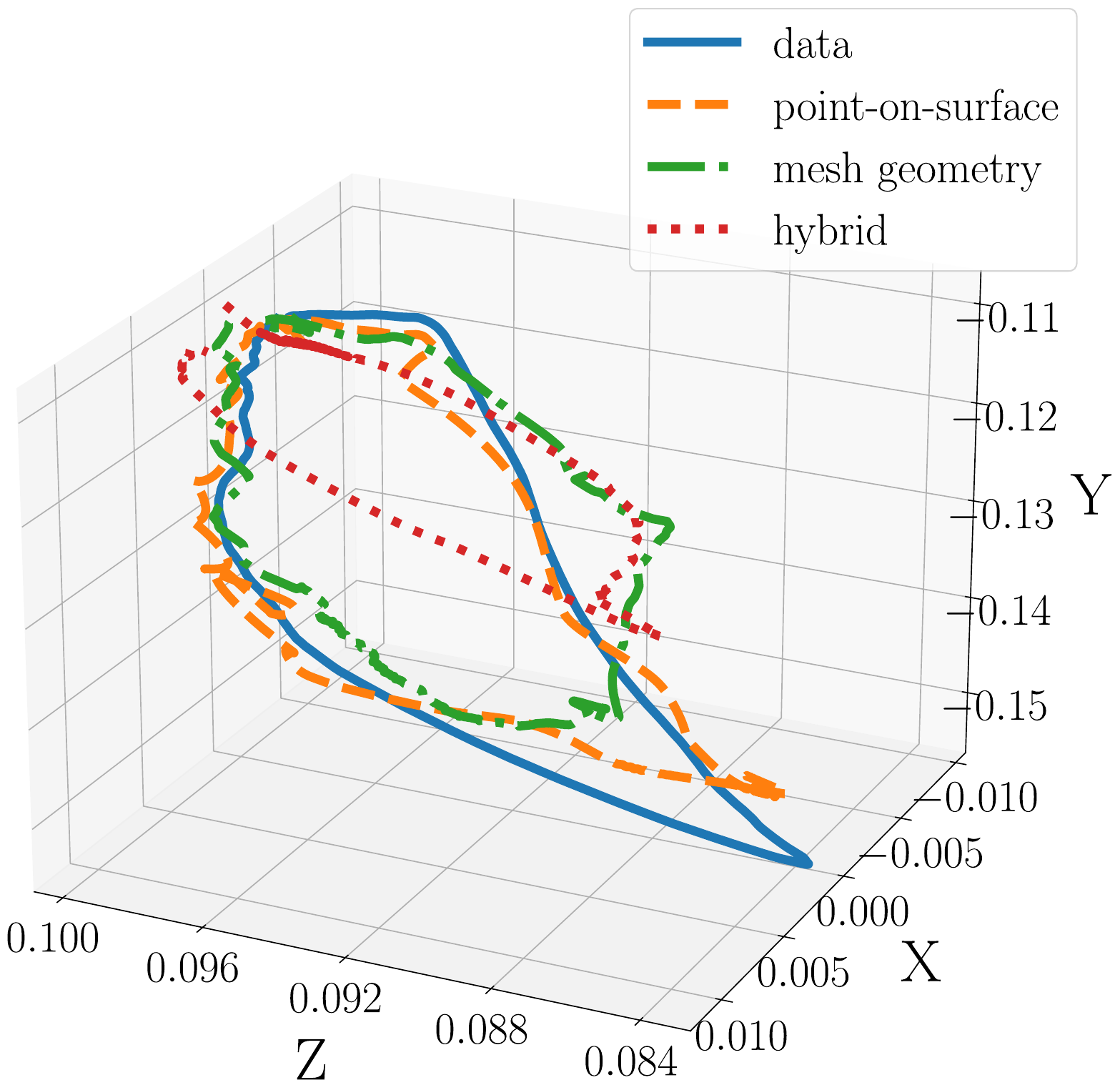}
    \captionsetup{font=small, labelfont=bf}
    \caption{A 3D plot of the recorded and simulated jaw movements with all model variants---point-on-plane constrained, mesh geometry contact constrained, hybrid---during chewing. The reference point on the jaw was the incisal point in the middle of the lower teeth. The unit is meters.}
    \label{fig:chewing-3d-plot}
\end{figure}

The results, summarized in \autoref{tab:performance-metrics}, demonstrate that the models can reproduce the recorded jaw movements. 
The average mean absolute error across all motions for the point-on-surface model was \SI{1.97}{mm}, for the mesh geometry model \SI{3.52}{mm}, and for the hybrid model \SI{6.35}{mm}. The performance is thus equal to or better than similar reported results in the literature \cite{Guo2022}.  
Note that the controller gains were only optimized for the point-on-surface constrained model and then later used for the other two models. Good performance was achieved with the point-on-surface constrained model, which produced the lowest errors across all metrics except the minimum distance error. Due to the loss of one degree of freedom in the point-on-surface constraint model, some locations are unreachable, resulting in a higher minimum distance error. Both the mesh geometry and hybrid models cannot reach the maximum opening distance due to the constraints imposed by the different anatomies of the participants and the models.   
The hybrid model shows the worst performance, especially for chewing, opening, and closing, since it contains the most uncertainty regarding model parameters and TMJ disc geometries \cite{Sagl2019a}. An exemplary chew cycle and the simulated positions of the mandible for all model variants can be seen in \autoref{fig:chewing-3d-plot}. The results highlight the importance of customizing the model to the specific patient to achieve more accurate kinematics and joint forces predictions, which will be a topic for future works.

\begin{table}[htbp]
    \caption{A performance comparison across the three model variants for different jaw motions (cyclic chewing motion, open/close, left/right, and protrusion/retrusion). The error metrics have the unit millimeters.}
    \begin{center}
    \begin{tabular}{p{1.05cm}|c|c|c|c|c}
        \textbf{motion} & \textbf{model variant} & \textbf{MAE}$^{\mathrm{a}}$ & \textbf{RMSE}$^{\mathrm{b}}$ & \textbf{Max}$^{\mathrm{c}}$ & \textbf{Min}$^{\mathrm{d}}$ \\
        \cline{1-6}
        \multirow{3}{*}{chewing} & point-on-surface & \num{2.08} & \num{2.49} & \num{6.72} & \num{0.51} \\
        & mesh geometry & \num{3.46} & \num{4.63} & \num{12.49} & \num{0.00} \\
        & hybrid & \num{9.01} & \num{11.03} & \num{22.66} & \num{1.63} \\
        \hline
        \multirow{3}{*}{open/close} & point-on-surface & \num{2.89} & \num{4.02} & \num{8.93} & \num{0.11} \\
        & mesh geometry & \num{4.89} & \num{7.02} & \num{16.34} & \num{0.00} \\
        & hybrid & \num{10.40} & \num{14.44} & \num{29.53} & \num{0.82} \\
        \hline
        \multirow{3}{*}{left/right} & point-on-surface & \num{1.56} & \num{1.71} & \num{2.89} & \num{0.22} \\
        & mesh geometry & \num{2.75} & \num{3.1} & \num{5.20} & \num{0.00} \\
        & hybrid & \num{2.91} & \num{3.02} & \num{4.82} & \num{0.1} \\
        \hline
        \multirow{3}{*}{protr./retr.} & point-on-surface & \num{1.33} & \num{1.45} & \num{2.74} & \num{0.51} \\
        & mesh geometry & \num{2.99} & \num{3.5} & \num{5.74} & \num{0.00} \\
        & hybrid & \num{3.09} & \num{3.76} & \num{6.18} & \num{0.01} \\
        \multicolumn{6}{l}{$^{\mathrm{a}}$mean absolute error (distance [mm]).} \\
        \multicolumn{6}{l}{$^{\mathrm{b}}$root mean squared error (distance [mm]).} \\
        \multicolumn{6}{l}{$^{\mathrm{c}}$maximum error [mm]. $^{\mathrm{d}}$minimum error [mm].}
    \end{tabular}
    \label{tab:performance-metrics}
    \end{center}
\end{table}%

\section{Discussion}\label{sec:discussion}

In the existing jaw model literature, muscle excitations are often manually tuned or optimized using static optimization methods to solve muscle recruitment problems. Dynamic optimization, however, remains challenging due to the high dimensionality of the problem. Although Guo \etal{} employed a forward-inverse dynamics procedure applying feedback control with PD controllers to control jaw movements, their approach did not involve optimization of the controller gains. Their work included seven participants but was limited to a single opening and closing motion \cite{Guo2022}. In contrast, the method demonstrated in this work produced good results for more complex jaw movements, such as chewing, while accounting for dynamics. However, the optimization is computationally expensive due to the use of a global optimization algorithm and the problem's dimensionality of \num{72}. A manually tuned PID controller may be sufficient for more straightforward tasks to analyze general jaw dynamics.

The basic rigid body model with point-on-surface constraints demonstrates promising results in reproducing recorded jaw movements. Consequently, it can effectively analyze general jaw dynamics and motions. However, this model is unsuitable for estimating joint forces, and an analytical and more computationally efficient model may be required for control tasks and controller design in the future. 
The evaluation of the second model highlights the critical importance of individualizing the model to specific TMD patients to achieve more accurate predictions of kinematics and joint forces. Nevertheless, its higher computational cost renders it impractical for control tasks or other real-time applications.  
The hybrid model might be the most valuable for estimating the effects of a jaw exoskeleton on the masticatory system and joints. However, its application is constrained by the lack of individualized data and noninvasive validation methods, which limit its use to qualitative assessments. By integrating rigid dynamics and FEM, the hybrid model strikes a balance between computational cost and accuracy. More basic models, meanwhile, remain effective for providing approximate muscle excitation estimates, which can serve as inputs for the FEM model when analyzing exoskeleton behavior and masticatory system functions. Nonetheless, further improvements in the hybrid model's performance must be explored and seem feasible as the model by Sagl \etal{} with a five times faster simulation time demonstrates \cite{Sagl2019a}.

The value and contribution to autonomous jaw robotics and exoskeletons are manifold. The presented framework serves as a research basis that can be used to develop and validate exoskeleton designs while considering ethical and user safety aspects, eliminating the need for human trials. The basic jaw models enable fast motion analysis and controller design testing. In contrast, the more detailed models allow for evaluating the effects of an exoskeleton on the masticatory system and joints, thus preventing potential harm to users during rehabilitation.
Furthermore, the framework can be directly applied to provide neurological and visual feedback to users for trajectory-tracking training or to visualize motions, which can be achieved using measured electromyography or kinematic data from the exoskeleton as model inputs. The possible individualization of the models enhances the accuracy of predictions and allows for personalized progress tracking. However, although model customization is made more convenient through the use of \textit{JSON} files, assembling the model components remains time-consuming, requiring the manual specification of all components and their attachment points.

Existing jaw exoskeleton systems are limited in their use of complex, validated jaw models as a basis for development and validation. This fact has raised ethical concerns, as well as concerns about user safety and the effectiveness of these exoskeletons in terms of power transmission and maximum joint loads. An advantage of the presented open-source framework is that it provides a common, shared foundation for further research. This benefit enables jaw exoskeleton researchers to focus directly on exoskeleton design without investing excessive time in jaw model development and evaluation. 

Although the presented jaw models can be further refined by incorporating more detailed representations of the hyoid bone, TMJ capsules, or cartilage, the focus of this work is not to deliver a perfect jaw model. Instead, the aim is to provide a foundational framework for developing jaw exoskeletons, which remains absent in the current literature. By including models with varying complexity levels and leveraging the framework's open-source nature, along with the abstraction of model components and properties, this framework is well-suited for prototyping and studying the behavior of exoskeletons designed for TMD treatment. It ensures a shared foundation for further research and innovation in this emerging field. Still, the framework does not yet include exoskeleton design and control tools, diminishing the usability for non-experts and the advantage gained by the abstraction of model components and properties.

\section{Conclusion}\label{sec:conclusion}

An extendable jaw model framework for the development of jaw exoskeletons aimed at the rehabilitation of TMDs has been proposed in this work. Three model variants, each with different levels of complexity and detail, were introduced. The evaluation was conducted by capturing kinematic data of the jaw using an optical motion capture system. Muscle excitations required to reproduce the recorded jaw movements were generated using PID controllers acting on each muscle, with controller parameters tuned through a global optimization scheme. The results demonstrated that the models can reproduce a real individual's jaw movements, even without prior personalization of the model. Still, individualization is necessary for more accurate predictions tailored to a specific TMD patient.
The basic rigid body models are suitable for rapid prototyping and studying jaw exoskeletons' general behavior. In contrast, the hybrid model might be the most valuable for estimating the effects of a jaw exoskeleton on the masticatory system and joints. The open-source nature of the framework ensures that researchers can collaborate and build upon a shared foundation, facilitating progress in this underexplored area.

Future work will focus on several key aspects, including the incorporation of more detailed model components; evaluation with a larger cohort of participants; the individualization of the model to enable more accurate predictions for specific TMD patients; the inclusion of more accurate FEM joint disc materials; the (automatic) creation of models derived from MRI data; the optimization of model parameters, such as maximum muscle force; and the development of tools for exoskeleton design and control.

This framework establishes a foundation for the development of jaw exoskeletons for the rehabilitation of TMDs, a research field that is still in its early stages.

\section*{Acknowledgment}

Optimization calculations for this research were conducted on the Lichtenberg high-performance computer of the TU Darmstadt.

\thanks{...
  \newline DOI: 10.1109/SIMPAR62925.2025.10979121
}

\end{document}